\documentclass[a4paper,12pt]{article}
\usepackage{amscd,amsmath,amssymb,amsthm,enumerate,indentfirst,longtable}
\usepackage{physics}
\usepackage{authblk}
\usepackage{mathptmx}
\usepackage{caption}
\usepackage[normalem]{ulem}
\usepackage[unicode]{hyperref}
\usepackage{csquotes}

\usepackage{graphics}

\usepackage{algorithm}
\usepackage[indLines=true,italicComments=true,rightComments=false]{algpseudocodex}
\algrenewcommand\algorithmicrequire{\textbf{Input:}}
\algrenewcommand\algorithmicensure{\textbf{Output:}}
\usepackage{afterpage}
\usepackage{multirow}
\usepackage{lastpage}
\usepackage{totcount}
\usepackage{listings}
\usepackage{threeparttable}
\usepackage{bbold}

\newcommand{\ZZ}{\mathbb{Z}}

\newcommand{\scalarprod}[1]{\left\langle #1\right\rangle}

\theoremstyle{definition}

\usepackage{fancyhdr}
\patchcmd{\chapter}{plain}{empty}{}{}

\renewcommand{\@oddhead}{\ifnum\thepage>1{\hfil\large\thepage\hfil}\fi}
\renewcommand{\@evenhead}{\ifnum\thepage>1{\hfil\large\thepage\hfil}\fi}
\title{Pitfalls of the sublinear QAOA-based \\ factorization algorithm}

\author[1,2]{Sergey V. Grebnev}
\author[1,2]{Maxim A. Gavreev}
\author[1,2]{Evgeniy O. Kiktenko}
\author[1,2]{Anton P. Guglya}
\author[2,3]{Albert R. Efimov}%
\author[1,2]{Aleksey K. Fedorov}

\affil[1]{Russian Quantum Center, Skolkovo, Moscow 121205, Russia}
\affil[2]{National University of Science and Technology “MISIS”, Moscow 119049, Russia}
\affil[3]{Sberbank of Russia, Sber Innovation and Research, Moscow 121357, Russia}

\begin{document}

\date{\today}

\maketitle

\begin{abstract}
Quantum computing devices are believed to be powerful in solving the prime factorization problem, which is at the heart of widely deployed public-key cryptographic tools.
However, the implementation of Shor's quantum factorization algorithm requires significant resources scaling linearly with the number size; 
taking into account an overhead that is required for quantum error correction the estimation is that 20 millions of (noisy) physical qubits are required for factoring 2048-bit RSA key in 8 hours. 
Recent proposal by Yan et. al.
claims a possibility of solving the factorization problem with sublinear quantum resources.
As we demonstrate in our work, this proposal lacks systematic analysis of the computational complexity of the classical part of the algorithm, which exploits the Schnorr's lattice-based approach. 
We provide several examples illustrating the need in additional resource analysis for the proposed quantum factorization algorithm.
\end{abstract}

\large

\section{Introduction}
\label{sec:introduction}

Most of the currently used public-key cryptosystems are believed to be secure against attacks with powerful conventional computers, but are not resistant to attacks with quantum computers~\cite{Brassard1998,Ladd2010,Fedorov2022}.
For instance, Shor's algorithm allows one to solve the integer factorization problem in polynomial time~\cite{Shor1994,Shor1999}, 
thus opening up prospects for efficient cryptoanalysis of existing public-key protocols.
Proof-of-concept experimental factoring of small integers such as 15, 21, and 35 have been successfully demonstrated on superconducting~\cite{Martinis2012-2}, trapped ion~\cite{Blatt2016}, and photonic~\cite{Pan2007,White2007,OBrien2012} quantum computers. 
One should mention, however, that resources required for realization of the Shor's algorithm for realistic key sizes are quite significant since the number of qubits (basic elements of quantum computing devices) 
scales linearly with the key sizes; thus, in order to recover a 2048-bit RSA secret key one needs more than four thousands of qubits, while the largest-scale available quantum computers only have about several hundreds of qubits~\cite{Lukin2022,IBMroadmap}.
Moreover, existing quantum computers belong to the class of noisy intermediate-scale quantum (NISQ) devices, whose workflow is highly affected by noise. 
Thus in order to run Shor's algorithm it is essential to improve the quality of operations under qubits (so-called gate fidelities) in order to make it possible to use quantum error correction codes.
Combining these facts altogether, the most realistic estimation as for today is that 20 millions of (noisy) physical qubits are required to recover a 2048-bit RSA key in 8 hours~\cite{Gidney2021}. 
Another recent proposal~\cite{Gouzien2021} suggests a way to factor 2048 RSA integers in 177 days with 13436 physical qubits and a multimode memory. 
A recent forecast review~\cite{Sevilla2020} estimates the likelihood for quantum devices capable of factoring RSA-2048 to evolve before 2039 as less than 5$\%$.

The considerable attention to the factorization problem using quantum resources has posed an interesting problem of reducing requirements to the quantum devices. 
As an example, there is an increasing interest in alternative schemes for solving the prime factorization problem using quantum tools, such as variational quantum factoring~\cite{Aspuru-Guzik2019,Karamlou2021};
however, they are generally not expected to perform significantly better than classical factoring algorithms.

Recent proposal by Yan et. al.~\cite{Yan2022} claims a possibility of solving the factorization problem with sublinear resources,
specifically, the number of qubits required for factoring a composite integer $N$ scales as $O(\log{N}/\log\log{N})$.
According to the estimation, a quantum circuit with 372 physical qubits only and a depth of thousands is necessary to challenge RSA-2048.  
The idea of the proposed method is to perform several steps  of the lattice reduction-based Schnorr's factorization algorithm~\cite{Schnorr2021} on a quantum computer in order to increase their rate of success.
The quantum part of the algorithm is based on quantum approximate optimization algorithm (QAOA)~\cite{Farhi2014,Farhi2019}, which can be efficiently run on NISQ devices~\cite{Monroe2020,Babbush2021,Bharti2021,Babbush2021-4}.
Taking into account the recent demonstrations of 433-qubit superconducting quantum processor~\cite{IBMroadmap} and 256-atom quantum simulator~\cite{Lukin2022},
quantum cryptanalysis of existing public-key cryptography schemes may be seen as the question of very near future. 
This circumstance caused intensive discussions in research blogs and open sources~\cite{Aaronson2023, Schneier2023, IDQ2023, QuantumCompRep2023, 2307.09651}.

In the present work, we demonstrate that the aforementioned proposal lacks systematic analysis of the computational complexity. 
We provide several examples illustrating both the inconsistencies in the description and the need in additional resource analysis for the aforementioned  factorization method.

\section{QAOA-based acceleration of Schnorr's factorization}

\subsection{Classical pre- and post-processing}

We restrict ourselves to the RSA case of the factorization problem: given an integer $N$ such that $N=p\cdot q$ for $p,q$ -- distinct primes close to $\sqrt{N}$, find  $p$ and $q$.
Let us introduce some basic definitions and notations first.
\begin{itemize}
    \item We denote $\mathbb{N}_+$ and $\mathbb{R}_+$ sets of positive integer and positive real numbers correspondingly.
    \item \emph{Factor base} $P_n$ is the set $\{p_i\}_{i=0,\ldots,n}$ of the first $n$ primes $(p_1=2, p_2=3, \ldots$) along with $p_0:=-1$.
    \item An integer $u$ is called \emph{$p_n$-smooth} if all of its prime factors are from $P_n$, so $u=\prod_{i=0}^n p_i^{e_i}$ for some non-negative integers $e_i$;
    \item For a given RSA-integer $N$ and factor base $P_n$, a pair of integers $(u,v)$ is called a \emph{useful relation} (\emph{a smooth relation pair}, \emph{sr-pair}) if $u$ and $u-vN$ are $p_n$-smooth.
    In particular, if $u>0$, then one has
    \begin{equation} \label{eq:u-and-v}
        u = \prod_{i=1}^n p_i^{e_i}, \quad u-vN=\prod_{i=0}^n p_i^{e_i'}
    \end{equation}
    for some nonnegative integers $e_i$ and $e_i'$.
    Note that for a sr-pair $(u,v)$ one also has
    \begin{equation} \label{eq:frac-equiv-1}
        \frac{u-vN}{u}\equiv 1\pmod{N}.
    \end{equation}
\end{itemize}

Next, we outline the basic steps of studied factorization algorithm~\cite{Schnorr2021,Yan2022}.
\begin{enumerate}
\item Construct a factor base $P_n$ for some fixed $n$.
In what follows we assume that $p_n\ll \sqrt{N}$, and so $N$ is not $p_n$-smooth.
Otherwise, $N$ can be factorized in ${\rm poly}(n)$ time.
\item Generate a set of $n+1$ useful relations $\{(u_j,v_j)\}_{j=1,\ldots,n+1}$ with respect to $P_n$. 
\item Construct matrices $E=(e_{i,j})$ and $E'=(e_{i,j}')$ with elements taken from factorization of $u_j$ and $v_j$ in $P_n$ correspondingly:
\begin{equation}
    u_j=\prod_{i=0}^{n} p_i^{e_{i,j}}, \quad 
    u_j-Nv_j=\prod_{i=0}^{n} p_i^{e_{i,j}'};
\end{equation}
\item Find solutions $\tau=(\tau_1, \ldots, \tau_{n+1})^\top\neq 0$ of the system of linear equations
\begin{equation} \label{eq:sleq}
    (E'-E)\begin{pmatrix}
        \tau_1 \\ \vdots \\ \tau_{n+1}
    \end{pmatrix} \equiv 0 ~({\rm mod}~2).
\end{equation}
and for each linear independent $\tau$ construct 
\begin{equation}
    X_\tau:=\prod_{i=0}^n p^{\frac{1}{2}\sum_{j=1}^m(e'_{i,j}-e_{i,j})\tau_j} ~\pmod{N}
\end{equation}
which definitely satisfies $X_\tau^2\equiv 1~({\rm mod}~N)$.
(The idea is that solutions of~\eqref{eq:sleq} provide even powers of primes in the expression
$\prod_{i=0}^n p^{\sum_{j=1}^m(e_{i,j}-e_{i,j}')\tau_j}\equiv 1\pmod{2}$, and allows taking a square root to obtain $X_\tau$.)
\item According to the classic Fermat's idea, if it appears that $X_\tau\not\equiv \pm 1~({\rm mod}~N)$, then factors of $N$ can be calculated efficiently as $\gcd(X_\tau\pm 1,N)$, where $\gcd$ stands for the greatest common divisor.
\end{enumerate}

We note that the most efficient known factorization method --- the number field sieve (NFS) --- 
uses a similar idea and reaches the complexity of $L_N[1/3,c]=\exp[(c+o(1))((\ln N)^{1/3}(\ln\ln N)^{2/3})]$ binary operations with $c\approx 1.903$ (see~\cite{lenstra_development_1993}).
This method was used to set the current factorization record for an RSA integer with 250 decimal digits, see Ref.~\cite{RSA250}.
The step of collecting the relations (the actual \emph{sieving}) dominates the complexity of the algorithm, thus, it is natural to search for new possibilities to obtain $N$-dependent smooth pairs.

In 2021, Schnorr proposed a lattice-based  approach to the procedure of relations collection~\cite{Schnorr2021}.
In order to find a relation, it was suggested to consider the solution ${\bf e}=\sum_{i=1}^{n+1}e_i{\bf b}_i$ of the \emph{shortest vector problem} (SVP) of a lattice $\Lambda\subset \ZZ^{n+1}$ defined by vectors
\begin{multline}
    \begin{pmatrix}
        {\bf b}_1 & \ldots & {\bf b}_{n+1}
    \end{pmatrix}:=\\
    \begin{pmatrix}
        f(1) & 0 & \dots & 0&0\\
        0 & f(2) & \dots & 0&0\\
        \vdots&\vdots&\ddots&\vdots&\vdots\\
        0 & 0 &\dots & f(n) & 0\\
        N'\ln p_1 & N'\ln p_2 & \ldots & N'\ln p_n & N'\ln N
    \end{pmatrix},
\end{multline}
where $N':=N^{1/(n+1)}$, $p_i $ is the $i$-th prime number, and $f:[1,\dots,n]\to [1,\dots,n]$ is a random permutation.
It is claimed~\cite{Schnorr2021} that $(u,v)$ defined as
\begin{equation} \label{eq:sr-pair-from-e}
    u:=\prod_{e_i> 0}p_i^{e_i}, \quad v:=\prod_{e_i< 0}p_i^{-e_i},
\end{equation}
form an sr-pair with high probability (here the first $n$ coordinates $e_1,\ldots, e_n$ are considered).
In this way, it was suggested to solve the factorization problem by solving a number of SVP problems.
Considering the efficiency of the algorithm, Schnorr claims that ``This [method] destroys the RSA cryptosystem''.

The main drawback of Schnorr's original algorithm is the lack of systematic complexity estimations. 
While the construction should work in general, no estimate for the required lattice dimension parameter $n$ as a function of $N$ were derived neither theoretically, 
nor empirically, while this parameter, in turn, defines the total complexity of both the sr-pair collection and the linear algebra parts of the algorithm. 
Moreover, a smaller $n$ may not provide enough ``useful'' permutations to gather enough linearly independent relations. For several counterexamples, see Ref.~\cite{Ducas}.
Notably, the same issues remain for the QAOA-enhanced version considered further.

In practice, there exist several approaches to construct a lattice  $\Lambda$ (see Refs.~\cite{Ducas,Yan2022,Schnorr2021}) by ``scaling'' the rightmost row in different manners. 
In Ref.~\cite{Yan2022} it was suggested to consider a closest vector problem (CVP) for lattices defined by $n$ $(n+1)$-dimensional vectors 
\begin{multline}
    \begin{pmatrix}
        {\bf b}_1 & \ldots & {\bf b}_{n}
    \end{pmatrix} :=\\
 \begin{pmatrix}
    \lceil f(1)/2 \rfloor & 0 & \ldots & 0 \\
    0 & \lceil f(2)/2 \rfloor & \ldots & 0\\
    \vdots & \vdots & \ddots & \vdots \\
        0&0&\ldots& \lceil f(n)/2 \rfloor \\
        \lceil 10^c \ln p_1  \rfloor & \lceil 10^c \ln p_2  \rfloor & \ldots & \lceil 10^c \ln p_n  \rfloor
\end{pmatrix}
\end{multline}
and the target vector
\begin{equation}
    {\bf t} = \begin{pmatrix}
        0 & \ldots & 0 & \lceil 10^c \ln N \rfloor
    \end{pmatrix}^\top
\end{equation}
where $\lceil \cdot \rfloor$ denotes rounding operation and a floating number $c>0$ is called a \emph{rounding parameter}.
The search of sr-pairs $(u_j,v_j)$ was suggested to perform from solutions of SVP problem ${\bf e}=\sum_{i=1}^{n}e_i{\bf b}_i$ in the same form~\eqref{eq:sr-pair-from-e}.

The main feature of the scheme considered in Ref.~\cite{Yan2022} is that SVP problems are solved in two steps.
First, Babai's CVP algorithm is applied to obtain an approximate solution
\begin{equation}
    {\bf b}_{\rm op}=\sum_{i=1}^n c_i {\bf d}_i,
\end{equation}
where ${\bf d}_1,\ldots,{\bf d}_n$ are vectors of LLL-reduced basis, and $c_i:=\lceil \mu_i \rfloor\in \mathbb{Z}$ are obtained by rounding real values $\mu_i\in\mathbb{R}$ (see Algorithm~\ref{alg:Babai} for details).
Then a QAOA is used to improve the solution in the form
\begin{equation}
    {\bf v}_{\rm new}=\sum_{i=1}^n (c_i+x_i){\bf d}_i,
\end{equation}
where $x_i\in\{-1,0\}$ if $c_i>\mu_i$ and $x_i\in\{0,1\}$ if $c_i\leq \mu_i$ (the QAOA is discussed in Sec.~\ref{sec:QAOA}).
In this way, the corrected solution is being searched in a ``unit neighborhood'' of Babai's solution: each integer coordinate in LLL-reduced basis is taken from $\{\lfloor \mu_i \rfloor, \lfloor \mu_i+1 \rfloor \}$.
Note that the workflow of Babai's algorithm is also specified by Boolean parameter ${\sf sort}$ indicating sorting of LLL-reduced basis vectors.

Another important modification of the considered algorithm compared to previous schemes is that it was also suggested to consider a larger factor bases $P_{B_2}$ with $B_2={\rm poly}(n)>n$ for checking smoothness of $u$ and $v$ obtained from SVP solutions (obtained from $(n+1)$-dimensional lattice).
We note that on the hand this trick increases the probability that given $(u,v)$ provides an sr-pair, but on the other hand, increases a number of equations for the system of linear equations of the form~\eqref{eq:sleq}.
The resulting workflow is presented in Algorithm~\ref{alg:Schnorr}.

\begin{algorithm}\caption{Babai's CVP algorithm ${\sf Babai}_{\delta,{\sf sort}}$}\label{alg:Babai} 
%\footnote{$n$ also defines the dimension of the lattice problems and the number of qubits, as we will see later}
\begin{algorithmic}[1]
 \Require 
$\mathbf{b}_1,\dots,\mathbf{b}_n\in\ZZ^{n+1}$\;\Comment{Basis vectors of the considered lattice $\Lambda$}\\
 ${\bf t}\in\ZZ^{n+1}$\; \Comment{Target vector}\\
 $\delta\in(0.25,1)$ [$\delta=3/4$ by default] \;\Comment{Parameter of the LLL-reduction}\\
  ${\sf sort}\in\{{\rm True},{\rm False}\}$ [${\sf sort}={\rm False}$ by default]\;\;\Comment{Rearranging parameter}    
 % A lattice $\Lambda$ given by a basis $ \mathbf{b}_1,\dots,\mathbf{b}_n\in\ZZ^{m}$, a target vector $\mathbf{t}\in\ZZ^m$
\Ensure 
${\bf c}=(c_1,\ldots,c_n)\in \mathbb{Z}^n$\;\Comment{Rounded coefficient of the solution in LLL-reduced basis}\\
${\bf \mu}=(\mu_1,\ldots,\mu_n)\in \mathbb{R}^n$\;\Comment{Unrounded coefficient of the solution in LLL-reduced basis}\\
$({\bf d}_1,\ldots,{\bf d}_n)$ \; \Comment{LLL-reduced basis}
%A vector $\mathbf{x}\in\Lambda$ s.t. $\parallel\mathbf{x}-\mathbf{t}\parallel\leq 2^{n/2}\mathrm{dist}(\mathbf{t}, \Lambda)$
\State ${\bf d}_1,\ldots, {\bf d}_n \leftarrow {\sf LLL}_{\delta}({\bf b}_1,\ldots,{\bf b}_n)$ \Comment{Constructing LLL-reduced basis via standard procedure}
\If{{\sf sort}}
    \State Rearrange $\{\mathbf{d}_i\}$ by ascension of their Euclidean norm
\EndIf
\State $(\widetilde{\bf d}_1,\ldots,\widetilde{\bf d}_n) \leftarrow $ Gram-Schmidt orthogonalization of $({\bf d}_1,\ldots,{\bf d}_n)$
\State $\mathbf{b}\leftarrow\mathbf{t}$
\For{$j = n, n-1, \ldots, 1$}
    \State $\mu_i \leftarrow \scalarprod{\mathbf{b},\widetilde{\mathbf{d}_j}}/\scalarprod{\widetilde{\mathbf{d}}_j,\widetilde{\mathbf{d}}_j}$
    \State $c_j \leftarrow \lceil \mu_j\rfloor$
    \State $\mathbf{b}\leftarrow \mathbf{b} - c_j\mathbf{d}_j$
\EndFor
\State \Return ${\bf c}, \boldsymbol{\mu}, ({\bf d}_1,\ldots,{\bf d}_n)$
\end{algorithmic}
\end{algorithm}

\begin{algorithm}\caption{QAOA-enhanced factorization algorithm}
\label{alg:Schnorr} 
\begin{algorithmic}[1]
    \Require 
    %RSA integer $N\in\mathbb{N}_+$ to be factorized,
    %number of qubits $n\in\mathbb{N}_+$ used in the QAOA
    $N\in\mathbb{N}_+$\; \Comment{RSA integer to be factorized}\;\\
    $n\in\mathbb{N}_+$\;\Comment{Number of qubits used in the QAOA}\\
    $B_2\in\mathbb{N}_+$ (s.t. $B_2\geq n$, $p_{B_2}<N$)\; \Comment{Size of a factor base used for checking sr-pairs }\\
    $c\in \mathbb{R}_+$\;\Comment{Rounding parameter}\\
    $\delta\in(0.25,1)$ [$\delta=3/4$ by default] \;\Comment{Parameter of the LLL-reduction in Babai's algorithm}\\
    ${\sf sort}\in\{{\rm True},{\rm False}\}$ [${\sf sort}={\rm False}$ by default]\;\;\Comment{Rearranging parameter for Babai's algorithm}    
\Ensure $p,q\in\mathbb{N}$ {\bf or} {\bf Fail} \Comment{Factors of $N$}

\State $i \leftarrow 0$ \Comment{Initialize a counter for found relations}
\While{$i<n+1$} \Comment{Accumulating sr-pairs}
    \State Fix a random permutation $f:[1,\dots,n]\mapsto [f(1),\dots,f(n)]$
    \For{$j=1,2,\ldots,n$} \Comment{Setting lattice's basis vectors}
        \State ${\bf b}_j[k]\leftarrow\begin{cases}
            \lceil f(k)/2 \rfloor & k=j \\
            0 & k\neq j \\
        \end{cases} \quad \text{~for~} k=1,2,\ldots,n$
        \State ${\bf b}_j[n+1]\leftarrow \lceil 10^c\ln p_k \rfloor$
    \EndFor
    \State ${\bf t}[k]\leftarrow 0 \quad \text{~for~}k=1,2,\ldots,n, \quad {\bf t}[n+1]\leftarrow\lceil 10^c\ln N \rfloor$ \Comment{Setting the target vector}
    %\State ${\bf d}_1,\ldots, {\bf d}_n \leftarrow {\sf LLL}_{\delta}({\bf b}_1,\ldots,{\bf b}_n)$ \Comment{Constructing LLL-reduced basis}
    \State ${\bf c}, \boldsymbol{\mu}, ({\bf d}_1,\ldots,{\bf d}_n) \leftarrow {\sf Babai}_{\delta,{\sf sort}}({\bf b}_1,\ldots,{\bf b}_n;{\bf t})$ \Comment{Applying Babai's algorithm to obtain rounded (${\bf c}$) and unrounded ($\boldsymbol{\mu}$) coefficients of solution in the LLL-reduced basis ${\bf d}_1,\ldots,{\bf d}_n$}
    \State ${\bf x}\leftarrow {\sf QAOA}({\bf c};\boldsymbol{\mu};{\bf d}_1,\ldots,{\bf d}_n;{\bf t})$ \Comment{Applying the QAOA to obtain corrections $x_i\in\{\pm1,0\}$}
    \State ${\bf v}_{\rm new}\leftarrow \sum_{j=1}^{n}(c_j+x_j){\bf d}_j$ \Comment{Updating CVP colution}
    \State ${\bf v}_{\rm new}=\sum_{j=1}^n e_j {\bf b}_j$ \Comment{Rewriting updated CVP solution it in the initial basis}
    \State $u\leftarrow\prod_{e_j> 0}p_j^{e_j},\quad v\leftarrow\prod_{e_j< 0}p_i^{-e_j}$ \Comment{Constructing candidates for an sr-pair}
    \If{$u-v\cdot N$ is $P_{B_2}$-smooth {\bf and} $(u,v)\not\in\{(u_j,v_j)\}_{j=1}^i$} \Comment{A new sr-pair has been found}
        \State $u_i\leftarrow u,\quad v_i\leftarrow v$ \Comment{Adding the sr-pair}
        \State $u_i=\prod_{j=0}^{B_2}p_j^{e_{j,i}} \quad  u_i-Nv_i=\prod_{j=0}^{B_2} p_j^{e_{j,i}'}$ \Comment{Decomposing $u_i$ and $u_i-Nv_i$ in $P_{B_2}$}
        \State $i\leftarrow i+1$ \Comment{Increasing counter and proceed to the next permutation}
    \EndIf
\EndWhile
\State $a_{i,j}\leftarrow (e'_{i,j}-e_{i,j})~{\rm mod}~2$ \Comment{Constructing the the system of linear equations}
\State ${\rm Ker}[(a_{i,j})]={\rm Span}(\tau_1,\ldots,\tau_m)$ \Comment{Extracting linearly independent solutions}
\For{$\tau=\tau_1,\ldots,\tau_m$} \Comment{Looking through the solutions}
    \State $X_\tau\leftarrow\prod_{i=0}^{B_2} p^{\frac{1}{2}\sum_{j=1}^m(e'_{i,j}-e_{i,j})\tau_j}~\pmod{N}$ \Comment{Constructing $X_\tau$ satisfying $X_\tau^2=1\pmod{N}$}

    \If{$X_{\tau} \not\equiv \pm 1\pmod{ N}$} \Comment{Success!}
        \State \Return {$\gcd(X_{\tau}+1,N), \gcd(X_{\tau}-1, N)$}
    \EndIf    
\EndFor
\State \Return {\textbf{Fail}} \Comment{No solutions found}
\end{algorithmic}
\end{algorithm}

\subsection{QAOA-based acceleration} \label{sec:QAOA}

Recall that the aim the QAOA is to improve the Babai's algorithm solution ${\bf b}_{\rm op}=\sum_{i=1}^n c_i {\bf d}_i$ 
by searching over vectors ${\bf b}_{\rm op}=\sum_{i=1}^n (c_i+x_i) {\bf d}_i$ with $x_i\in\{-1,0\}$ or $x_i\in\{0,1\}$ depending on the sign of each $c_i-\mu_i$.
To employ the QAOA we first need to embed our cost function into a qubit Hamiltonian $\hat{H}_{\rm c}$ -- an energy observable over a system of spin-1/2 particles.
Consider the CVP cost function written with respect to ${\bf x}=(x_1,\ldots,x_n)$ of the form
\begin{multline}
    F({\bf x}) = \| {\bf v}_{\rm new} - {\bf t} \|^2=
    \|\sum_{i=1}^n (c_i+x_i) {\bf d}_i-{\bf t}\|\\=
    \sum_{j=1}^{n+1}\left((c_i+x_i) ({\bf d}_i)_j-({\bf t})_j\right)^2,
\end{multline}
where $\|\cdot\|$ is a standard Euclidean norm, ${\bf t}$ is the target vector, and $(\cdot)_j$ denotes $j$-th component of an element from $\ZZ^{n+1}$.
The QAOA ``target'' Hamiltonian $\hat{H}_{\rm c}$ is obtained from $F({\bf x})$ by replacing each $x_i$ with an $n$-qubit operator according to the following transformation:
\begin{equation} \label{eq:encoding}
x_{i} \rightarrow \hat{x}_i=     
    \begin{cases}
      \frac{1}{2}(\hat{\mathbb{1}} - \hat{\sigma}_{\rm z}^{i}), &\text{if~} c_{i} \leq \mu_{i} \\
      \frac{1}{2}(\hat{\sigma}_{\rm z}^{i} - \hat{\mathbb{1}}), &\text{if~} c_{i} > \mu_{i}\\
    \end{cases},
\end{equation}
where $\hat{\sigma}_z^i=\hat{1}\otimes\ldots\otimes\hat{1}\otimes\hat{\sigma}_z\otimes\hat{1}\otimes\ldots\otimes\hat{1}$ (here $\hat{\sigma}_z$ stands in $i$-th position of the tensor product of $n$ operators), $\sigma_z=\ket{0}\bra{0}-\ket{1}\bra{1}$ and $\hat{1}=\ket{0}\bra{0}+\ket{1}\bra{1}$ are standard single-qubit Pauli-$Z$ and identity operators correspondingly, $\ket{0}$ and $\ket{1}$ denotes single-qubit computational basis,  and $\hat{\mathbb{1}}\equiv\hat{1}^{\otimes n}$.
Note that eigenvalues of $\hat{x}_i$ coincide with possible values of $x_i$.
In the result we obtain an Ising-type Hamiltonian
\begin{equation}\label{Hamiltonian}
    \hat{H}_{\rm c} = \sum_{i=1}^n h_i \hat{\sigma}_z^i + \sum_{i,j=1}^n J_{i,j} \hat{\sigma}_z^i \hat{\sigma}_z^j,
\end{equation}
with real coefficients $\{h_i\}$ and $\{J_{i,j}\}$ calculated from $\{({\bf d}_i)_j\}, \{({\bf t})_j\}$, and $\{c_i\}$.

If one can prepare an $n$-qubit state $\ket{\psi}$ that minimizes the corresponding mean energy $\bra{\psi}H_{\rm c}\ket{\psi}$, then one can obtain a solution of the initial problem of minimizing $F(\bf x)$.
The solution vector ${\bf x}$ can be obtained by measuring $\hat{\sigma}_z$ for all qubits, thus getting $\sigma_z^i\in\{\pm1\}$ for each $i=1,\ldots,n$, and
then assigning every $x_i$ with $(1-\sigma_z^i)/2$ or $(\sigma_z^i-1)/2$ depending on whether $c_i\leq \mu_i$ or $c_i>\mu_i$ respectively.

The idea behind the QAOA is to approximate the optimal state $\ket{\psi}$ by a state prepared with parametrized quantum circuit with a layered  specific structure.
These circuits allow preparing $n$-qubit states of the form
\begin{equation}
    \ket{\gamma, \beta} = e^{-i\beta_{p} \hat{H}_{\rm mix}}e^{-i\gamma_{p} \hat{H}_{\rm c}} \dots e^{-i\beta_{1} \hat{H}_{\rm mix}}e^{-i\gamma_{1} \hat{H}_{\rm c}} \ket{+}^{\otimes n},
\end{equation}
where $\gamma=(\gamma_1,\ldots,\gamma_p)$, $\beta=(\beta_1,\ldots,\beta_p)$ with $\gamma_i,\beta_i\in \mathbb{R}$ are parameters of the quantum circuit used for preparing the state, $p\in\mathbb{N}_+$ is number of layers in the circuit, $\hat{H}_{\rm mix} = \sum_{j=1}^{n} \hat{\sigma}_{\rm x}^{j}$  
is the ``mixing'' Hamiltonian, which is a sum of Pauli-X $\hat{\sigma}_x=\ket{0}\bra{1}+\ket{1}\bra{0}$ operators acting of each qubit, and
$\ket{+}\equiv 2^{-1/2}(\ket{0}+\ket{1})$ (note that $\hat{\sigma}_x\ket{+}=\ket{+}$).
Quantum processors allow efficiently preparing states $\ket{\gamma, \beta}$ with circuits consisting of single- and two-qubit gates and having depth proportional to number of layer $p$. 
The cost function value $E(\gamma,\beta)=\bra{\gamma,\beta}H_{\rm c}\ket{\gamma,\beta}$ can be easily calculated from results of repeated computational basis measurement of $n$-qubit register.
The optimal values of parameters $(\gamma,\beta)$, which minimize $E(\gamma,\beta)$, are searched by using classical optimization routine.
We note that existing quantum processors suffer from noises (increasing with $p$), and so the resulting measurement statistics for real processors differs from the one corresponding to ``ideal'' $\ket{\gamma, \beta}$, that complicates the optimization process.

The runtime of the QAOA is composed of two parts: (i) running circuits on a quantum hardware for getting estimates of 
$\bra{\gamma,\beta}H_{\rm c}\ket{\gamma,\beta}$ for given $(\gamma,\beta)$  with enough precision and getting the resulting outputs, and (ii) updating $(\gamma,\beta)$ with a classical optimization routine.
We note the number of runs (also known as ``shots'') of a fixed circuit, which affects the accuracy of $\bra{\gamma,\beta}H_{\rm c}\ket{\gamma,\beta}$ estimation, serves as a hyperparameter of the algorithm.
It affects the performance of the classical optimization that makes realistic complexity analysis extremely difficult.

The results of Ref.~\cite{Yan2022}, as well as our numerical simulations, shows that it is worth to consider several output stings ${\bf x}$, which provide cost-function values near the minimum, for constructing $(u,v)$ pairs for the same permutation $f$.
The reported results of Ref.~\cite{Yan2022}, as well as our numerical simulations, show that quite often one of values ${\bf x}$ corresponding to, e.g.,
ten lowest eigenvalues of $\hat{H}_{\rm c}$ provides an sr-pair even though the optimal ${\bf x}$ (true ground state) does not.

\section{Analysis of the algorithm}

In order to analyze the proposed factorization algorithm, we reproduced all the intermediate steps for three RSA integers and corresponding hyperparameters ($n, B_2, c$) considered in Ref.~\cite{Yan2022} (see Table~\ref{tab:targets}).
We made our own implementation of Algorithms~\ref{alg:Schnorr} and~\ref{alg:Babai}, as well as the QAOA realized using an emulator of an ideal quantum processor, run on a classical computer.
The considered number of qubits $n=3,5,10$ allows for a brute-force search over all ${\bf x}\in\{0,1\}^n$, so we also considered in idealized realization of the QAOA, which outputs  strings ${\bf x}$ corresponding to $K=10$ lowest eigenvalues of $\hat{H}_{\rm c}$.
Our implementation was made in \texttt{SageMath} (version 9.8) and \texttt{Python} (version 3.11) using the \texttt{qiskit} library (version 0.41).
We note that parameters $\delta$ and ${\sf sort}$ were included in Babai's algorithm just in order to represent numerical results of Ref.~\cite{Yan2022}, where it is declared that $\delta=3/4$ is used and some sorting of vectors from LLL-reduced basis is mentioned.

\begin{table*}[! htbp]
\caption{Sets of parameters considered in Ref.~\cite{Yan2022}.}
\label{tab:targets}
\centering  
\begin{tabular}{|c|c|c|c|}
 \hline
RSA integer $N$ & Qubit number $n$ ($B_1$) & Rounding parameter $c$ & Factor base size $B_2$ \\
  %& $N$ & parameter $\epsilon$ \\
\hline
1961 & 3 & 1.5 & 15 \\
\hline
48567227 & 5  & 4 & 50  \\
\hline
261980999226229 & 10 & 4 & 200 \\
\hline
\end{tabular}
\end{table*}

The results of reproducing LLL-reduced basis vectors (Eqs.~(S40-S42) in Ref.~\cite{Yan2022}), Babai's algorithm solutions (Eqs.~(S43-S45) in~\cite{Yan2022}), 
and $(u,v)$ pairs obtained from the QAOA (Tables (S6-S8) in~\cite{Yan2022}) for given permutations $f$ and target vectors (Eqs.~(S33-38) in~\cite{Yan2022})  are presented in Table~\ref{tab:reproduction-full}.
According to this results, it seems that the LLL-reduction was performed with $\delta=0.99$ (the default value for standard LLL-reduction procedure).
Notably, in the cases of $n=3$ and $5$, even if the Babai's algorithm provide a different solution with ($\delta=3/4, {\sf sort}={\rm False}$) 
compared to ($\delta=3/4, {\sf sort}={\rm True}$) and  ($\delta=0.99, {\sf sort}={\rm False}$), the resulting $(u,v)$ pairs came from the QAOA appeared to be the same.
In the case of $n=10$, there is already a difference between the obtained pairs, however, a number of pairs from the bottom energy levels of the QAOA [5 for ($\delta=3/4, {\sf sort}={\rm False}$) and 4 ($\delta=3/4, {\sf sort}={\rm True}$)] appeared exactly the same as for ($\delta=0.99, {\sf sort}={\rm False}$).
The remaining overlaps between pairs appear at higher levels.
All the sr-pairs come the QAOA solutions were obtained exactly same as in~\cite{Yan2022} for all considered parameters of $\delta$ and ${\sf sort}$.
In this way, given that the QAOA provides a number of bottom eigenvalues of $\hat{H}_{\rm c}$, we verified and sr-pairs indeed can be obtained.

\begin{table*}[]
    \caption{Reproduction of Ref.~\cite{Yan2022} results for three considered RSA integers and different parameters $\delta$ and ${\sf sort}$  of Babai's algorithm.
    $(u,v)$-pairs are obtained from the lowest energies of the corresponding QAOA Hamiltonian.}
    \label{tab:reproduction-full}
    \centering $N=1961$ ($n=3$)\\
    \begin{tabular}{|c|c|c|c|}\hline
        & $\delta=\frac{3}{4}, {\sf sort}={\rm F}$ & 
          $\delta=\frac{3}{4}, {\sf sort}={\rm T}$ &
          $\delta=0.99, {\sf sort}={\rm F}$\\\hline
        LLL-reduced basis & $\times$ & $\checkmark$ & $\checkmark$  \\ 
        Babai's algorithm solution & $\times$ & $\checkmark$ & $\checkmark$ \\
        $(u,v)$-pairs from the QAOA & 4 out of 4 & 4 out of 4 & 4 out of 4 \\
        Smooth pairs & 3 out of 3 & 3 out of 3 & 3 out of 3\\
        \hline
    \end{tabular}\\\vspace{10pt}
    $N=48567227$ ($n=5$)\\
    \begin{tabular}{|c|c|c|c|}\hline
        & $\delta=\frac{3}{4}, {\sf sort}={\rm F}$ & 
          $\delta=\frac{3}{4}, {\sf sort}={\rm T}$ &
          $\delta=0.99, {\sf sort}={\rm F}$\\\hline
        LLL-reduced basis & $\times$ & $\checkmark$ & $\checkmark$  \\ 
        Babai's algorithm solution & $\times$ & $\checkmark$ & $\checkmark$ \\
        $(u,v)$-pairs from the QAOA & 10 out of 10 & 10 out of 10 & 10 out of 10\\
        Smooth pairs & 1 out of 1 & 1 out of 1 & 1 out of 1\\\hline
    \end{tabular}\\ \vspace{10 pt}
    $N=261980999226229$ ($n=10$)\\
    %\begin{threeparttable}
    \begin{tabular}{|c|c|c|c|}\hline
        & $\delta=\frac{3}{4}, {\sf sort}={\rm F}$ & 
          $\delta=\frac{3}{4}, {\sf sort}={\rm T}$  &
          $\delta=0.99, {\sf sort}={\rm F}$\\\hline
        LLL-reduced basis & $\times$ & $\times$ & $\checkmark$  \\ 
        Babai's algorithm solution & $\times$ & $\times$ & $\checkmark$ \\
        $(u,v)$-pairs from the QAOA & 8 out of 10 & 6 out of 10 & 10 out of 10 \\
        Smooth pairs & 1 out of 1 & 1 out of 1 & 1 out of 1\\\hline
    \end{tabular}\\
\end{table*}

The problems appear when we considered the whole workflow of Algorithm~\ref{alg:Schnorr}.
Recall, that the idea is to accumulate a number of sr-pair in order to solve a system of linear equations~\eqref{eq:sleq}.
In Ref.~\cite{Yan2022}, obtaining of only a single sr-pair for each $N$ is considered. 
It is stated that ``The calculations of other sr-pairs are similar and will be obtained by numerical method'' without clarifications what a particular method should be used.
One may suggest that all necessary relations can be obtained from other permutations $f$, however, it turns out that this  may not the case.
Our simulations show that for the considered case of $N=1961$, $n=3$, $B_2=15$, only $9$ can be obtained from QAOA solutions for all possible permutations $n$.

\begin{figure}[h]
\centering
\includegraphics[width=1.\textwidth]{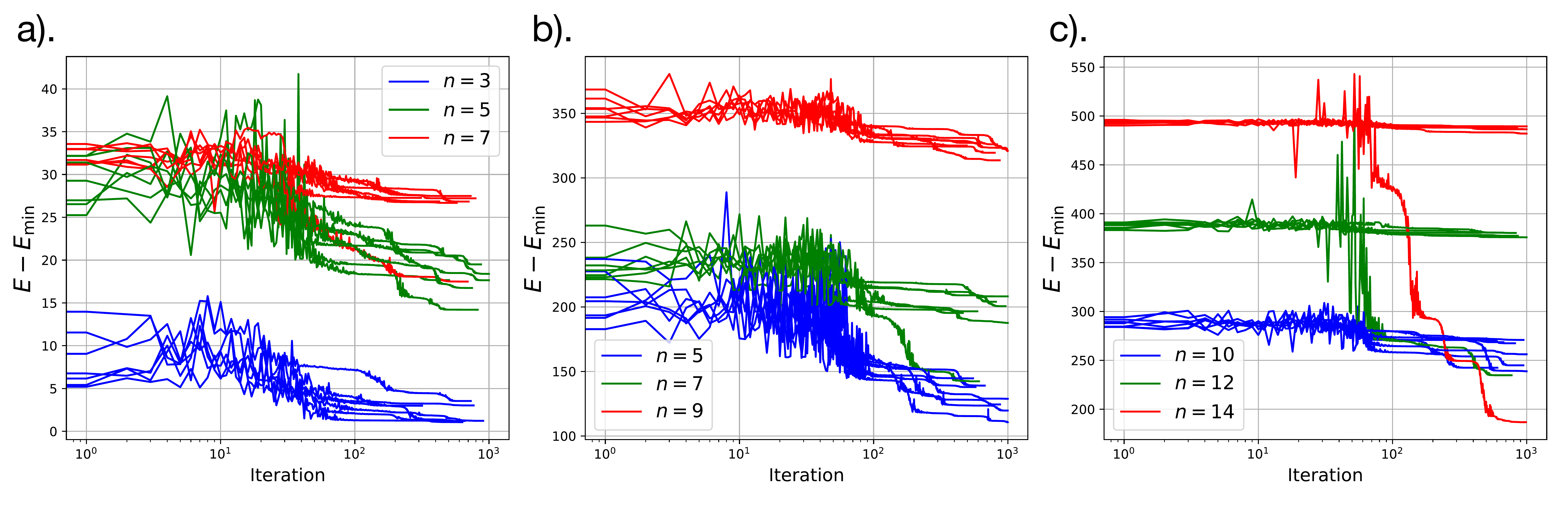}
\caption{{Convergence of mean energy $E$ to the minimal energy $E_{\min}$ for QAOA circuits consisting of $p=3$ 
layers designed for different number of qubits $n$ and different RSA integers $N=1961$~(a), $N=48567227$~(b), and $N=261980999226229$~(c). Each line corresponds to a distinct run of Nelder--Mead optimization algorithm.}}
\label{fig:test}
\end{figure}

In what follows, we formulate the main results of our analysis of Ref.~\cite{Yan2022}.
\begin{enumerate}
    \item Although it is shown that the QAOA provides sr-pairs for given permutations (a single one per each considered $N$), the validity of the whole factorization algorithm is not justified, since other sr-pairs are obtained using some unnamed ``numerical method''. 
    Our preliminary numerical results shows that at least for the considered $n=3$-qubit case, even going through all possible permutations does not provide the necessary number of relations.
    \item We found the statement that the proposed algorithm allows factoring integers with sublinear number of qubits rather confusing also due to a apparent scalability issues  related to both ``classical'' and ``quantum'' parts.
    \begin{itemize}
        \item On the one hand, the QAOA searches for an optimal vector among the closest neighbors of the quasi-shortest vector given by LLL-reduction. 
    The LLL algorithm gives a polynomial-time solution of the $\alpha$-SVP with $\alpha = (2/\sqrt{3})^n$, thus, increasing the dimension of the lattice, we exponentially decrease the probability of existence of such a vector.
        \item  On the other hand, there are no known results on the convergence of the QAOA for the problem of larger dimensions. 
        The results of our own simulation of the QAOA, shown in Fig.~\ref{fig:test}, shows convergence issues that intensify with a growth of $n$.
    \end{itemize}
    \item The workflow of the algorithm is governed by a number of hyperparameters such as $n$ (number of qubits), $B_2$ (size of a factor base used for checking relations), and $c$ (rounding parameter for construction of the lattice), whose appropriate choice is not specified for particular values of $N$.
    In particular, while testing a relation for smoothness, the authors of Ref.~\cite{Yan2022} arbitrarily use an extended factor base of $B_2=n^2$ primes, which naturally rises the smoothness probability of $u,v$, but leads to a quadratic growth of the size of the linear system and so a number of required relations.
    \item All the reported in Ref.~\cite{Yan2022} results seems to obtained by using LLL-reduced basis with $\delta=0.99$ rather than $\delta=3/4$.
    The formal description of algorithms in Ref.~\cite{Yan2022} posses some technical inaccuracies, e.g., the Gram-Schmidt orthogonalization step in Babai's algorithm is missing.
\end{enumerate}

\section{Conclusion}

We conclude that the results obtained in Refs.~\cite{Schnorr2021, Yan2022} in the current form are not sound enough to justify the terrific statements of both papers considering the resources required to break the actual instantiations of the RSA cryptosystem. 
We claim the lack of in-depth analysis of the complexity of both the classical and ``quantum-accelerated'' algorithms, as well as major inconsistencies in the description of the latter.
We also notice that, unlike the generic Shor's algorithm,  Schnorr's method does not apply to the discrete logarithm in a prime group setting, leaving the elliptic curve case alone, neither does it affect lattice-based post-quantum cryptography. 

We state, however, that the quantum acceleration approach may give rise to novel methods of cryptanalysis, which can be potentially easier in the implementation in comparison with the known methods. 
This consideration adds a serious reason to switch to quantum-secured tools, such as quantum key distribution~\cite{Gisin2001,Diamanti2016} and post-quantum cryptography~\cite{Bernstein2017,Fedorov2021-2}. 

\section*{Acknowledgment}
The research leading to these results was supported by Sberbank.
Part of this work was also supported by the RSF Grant \textnumero 19-71-10092. 

\section{Appendix A. QAOA simulation details}
To reproduce the results of the QAOA algorithm, we use the standard functionality of the \texttt{qiskit} library, which can simulate the behavior of the QAOA algorithm.
Namely, it provides the possibility to obtain exact probability distribution of read-out measurement outcomes, and thus to calculate the exact value of energy function $E(\gamma,\beta)=\bra{\gamma,\beta}H_{\rm c}\ket{\gamma,\beta}$ for given $\gamma$ and $\beta$.
To make an optimization over $(\gamma, \beta)$, we employ the Nelder-Mead optimization algorithm with parameters listed in Table~\ref{tab:NMpars} (set by default in the \texttt{qiskit} library).
We note that the energy estimation based on finite measurement statistics, which takes place if a real quantum processor is used, leads to a worse performance of the optimization algorithm compared to the one based on exact probability distributions.
In this way, the results shown in Fig. \ref{fig:test} can be considered as optimistic estimates of the real performance.
We also note that the dependence profiles of the cost function in Fig.~\ref{fig:test} indicates the presence of barren plateaus~\cite{BarrenPletaus} in the cost function landscape.

\begin{figure}[h]
\centering
\includegraphics[width=1.\textwidth]{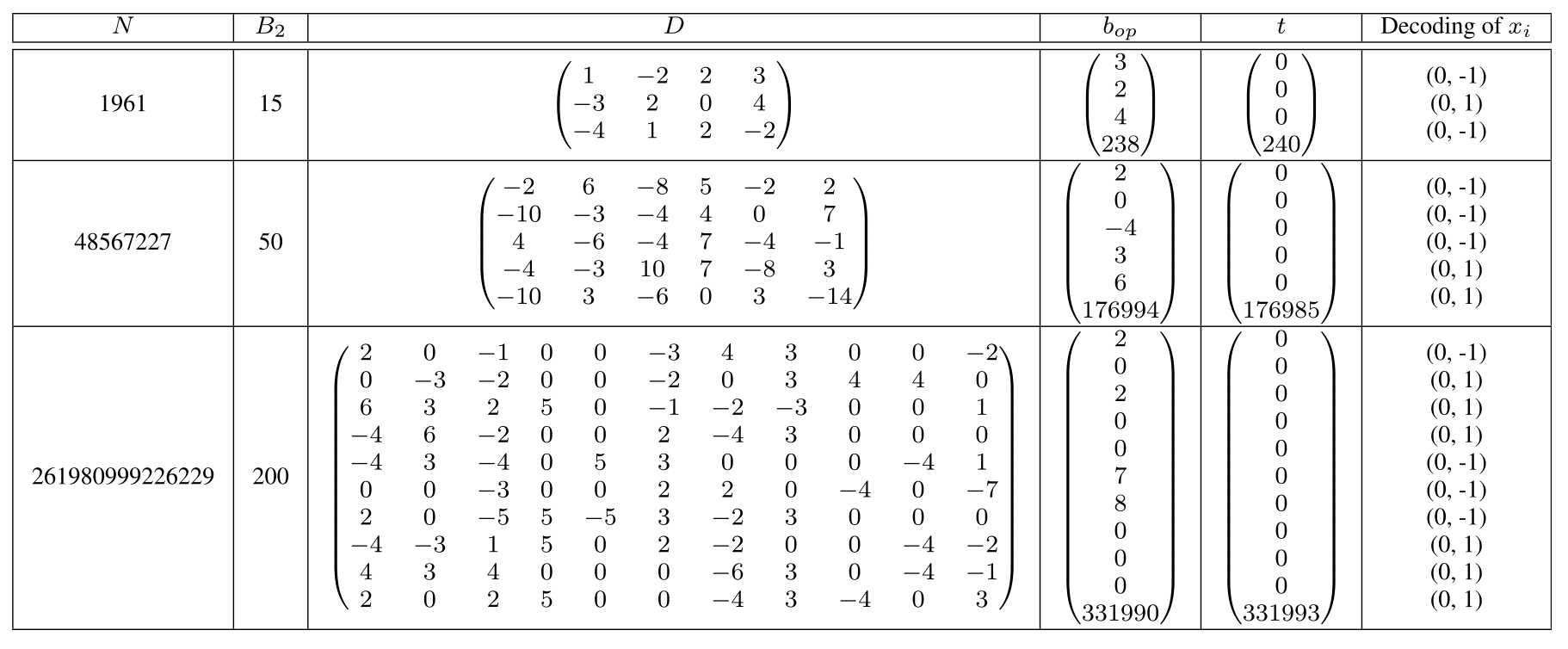}
\caption{{Numerical values of problem parameters for QAOA.
The rightmost column shows values $(x_i[\sigma^i_z=+1], x_i[\sigma^i_z=-1])$ of $x_i$ depending on measured value $\sigma^i_z$ for each $i=1,\ldots,n$, according to Eq.~\eqref{eq:encoding}.}}
\label{tab:numvals}
\end{figure}

To enable the reproduction of our results, below we provide the numerical values of problem parameters in Fig.~\ref{tab:numvals}.

\begin{table}[!th]
\caption{Parameters of the Nelder-Mead algorithm employed within the QAOA (default values set in the \texttt{qiskit} library).}
\label{tab:NMpars}
\centering  
\begin{tabular}{|c|c|} \hline
Parameter & Value \\ \hline
Maximum allowed number of iterations (maxiter) & None \\
Maximum allowed number of function evaluations (maxfev) & 1000 \\
Absolute tolerable error in the argument (xatol) & 0.0001 \\
Tolerance for termination (tol) & None \\
Adaptation to  dimensionality (adaptive) & False \\
\hline
\end{tabular}
\end{table}

\bibliographystyle{plain}
\bibliography{schnorr-v2}
\end{document}